\documentclass[%
 reprint,
 prd,
 amsmath,amssymb,
 aps,
 floatfix,
]{revtex4-2}

\usepackage{graphicx}
\usepackage{dcolumn}
\usepackage{bm}
\usepackage{hyperref}
\usepackage{array}
\usepackage{makecell}

\newcommand{\dd}{\mathrm{d}}

\newcommand{\E}{\mathbb{E}}
\newcommand{\Gvec}{\bm{G}}
\newcommand{\rhovec}{\bm{\rho}}
\newcommand{\Sig}{\bm{\Sigma}}
\newcommand{\Kmat}{\bm{K}}
\newcommand{\wpeak}{\omega_{\rm peak}}
\newcommand{\rpeak}{\rho_{\rm peak}}
\newcommand{\Wlow}{W_{\rm low}}
\newcommand{\Ntau}{N_{\tau}}

\begin{document}

\title{Conditional Model-Adequacy Tests for Spectral Uncertainty Claims in Lattice QCD}

\author{Haozheng Li}
 \altaffiliation[]{haozhengli2002@stu.pku.edu.cn}
\affiliation{%
 Center for High Energy Physics, Peking University, Beijing 100871, People's Republic of China
}%

\begin{abstract}
Euclidean lattice correlators determine spectral functions only through a smoothing integral transform, so a nominal uncertainty band on a reconstructed spectrum need not have a coverage interpretation for a physical summary.  We formulate this as a target-wise adequacy test for reported spectral uncertainties.  For a chosen summary \(T[\rho]\), the reported interval is tested on Euclidean-admissible mock correlators with known truth using empirical coverage, simulation-based calibration ranks, physical diagnostics, and stress tests.

The test is conditional, but it is a useful falsification tool: passing it does not prove that a reconstruction is the QCD truth, while failing it shows that the reported uncertainty law is not adequate for the chosen functional under the stated mock extension.  In a generic benchmark, peak locations are substantially better calibrated than peak heights or low-frequency weights, reflecting different degrees of functional identifiability under the Euclidean kernel.

We then apply the same logic to a finite-temperature shear correlator.  A family of BG-style reconstructions is compatible with the Euclidean data at \(\chi^2/N_\tau\simeq 1.3\).  Within the scanned grid and stated observable-matched mock extension, a \(W_{\rm low}\)-calibrated representative can be identified, whereas pointwise peak-height intervals are not certified for the tested BG-style uncertainty law.  Thus Euclidean compatibility is a necessary consistency check, but not a sufficient adequacy criterion for spectral uncertainty claims.
\end{abstract}

\maketitle

\section{Introduction}
\label{sec:introduction}

Lattice calculations give Euclidean correlators, whereas many real-time observables are encoded in spectral functions.  In the channels considered below the forward problem has the form
\begin{equation}
  G(\tau)=\int_0^\infty \dd\omega\, K(\tau,\omega)\rho(\omega),
  \label{eq:intro_forward}
\end{equation}
with a known kernel.  Transport is a canonical example: Kubo relations connect viscosities, conductivities, and diffusion coefficients to the low-frequency behavior of spectral functions \cite{Kubo:1957,Aarts:2007wj,Meyer:2007ic,Meyer:2008sn,Meyer:2011gj,Aarts:2021bqy,Astrakhantsev:2017nrs,Altenkort:2023eav}.  The inverse problem is ill conditioned because the kernel smooths the spectrum, the number of Euclidean time slices is small, and the data are noisy and correlated.  This has motivated maximum entropy methods, Bayesian reconstruction, Backus--Gilbert estimators, stochastic analytic continuation, Gaussian-process and spectral-density methods, and neural-network approaches \cite{Bryan:1990,Jarrell:1996rr,Nakahara:1999zkr,Asakawa:2000tr,Burnier:2013nla,Backus:1968zz,Sandvik:1998sac,Beach:2004sac,Hansen:2019idp,Bailas:2020qmv,DelDebbio:2025bg,Ding:2018zrk,Kades:2020zeg,Horak:2022gp,Wang:2022yqf,Buzzicotti:2024lm,Huang:2024spe}.

This paper addresses a different question from the choice of central spectral estimator.  A band obtained from an entropy-regularized local Gaussian approximation, a resolution-kernel construction, a bootstrap ensemble, a learned conditional sampler, or another uncertainty mechanism need not have the same repeated-experiment meaning \cite{Tierney:1986laplace,Efron:1993bootstrap,Rothkopf:2022fkc,Frison:2024bayeslqcd,Jay:202590}.  A visually plausible central spectrum, a small Euclidean residual, or a spread over ansatz choices does not by itself establish coverage of a nominal interval for a specified spectral functional.  Nevertheless, physical conclusions are often drawn from precisely such bands: whether a peak has moved, whether an amplitude is resolved, or whether a low-frequency sector is constrained enough to motivate a transport statement.  The operational question is therefore: under a specified Euclidean-admissible data-generating process, does a nominal interval contain the true value of the physical summary with the advertised frequency?

The answer is not expected to be universal because the Euclidean kernel does not constrain all spectral directions equally.  We use ``functional identifiability'' in an operational sense: the extent to which changes in a functional \(T[\rho]\) produce distinguishable changes in Euclidean correlators under the stated covariance and admissibility assumptions.  Moving a dominant peak usually changes the correlator coherently over many Euclidean times.  By contrast, an infrared integral, a local peak height, or a narrow shape deformation can be partially hidden by threshold motion, broadening, redistribution among nearby bins, positivity constraints, normalization constraints, or compensation against higher-frequency strength.  Thus an interval that is calibrated for \(\wpeak\) need not be calibrated for
\begin{equation}
  \Wlow(\omega_c)=\int_0^{\omega_c}\rho(\omega)\,\dd\omega,
  \label{eq:intro_wlow}
\end{equation}
or for \(\rpeak\).  The calibration problem is therefore functional dependent: Euclidean compatibility of a central reconstruction is necessary, but it does not imply calibrated uncertainty for every physical spectral functional.

This interpretation is asymmetric.  A successful mock calibration does not prove that the reconstructed spectrum is the QCD spectral function, because the mock ensemble is only a specified extension of the inverse problem.  A failure is nevertheless informative: if a reported uncertainty law cannot cover a target functional even in a Euclidean-compatible challenge ensemble, then that report should not be used to support the corresponding physical claim without additional assumptions.  The calibration test is therefore a conditional adequacy test for uncertainty claims, not a probability assignment to nature.

The purpose of this work is to turn the reliability of a reported spectral interval into a falsifiable, target-wise statement.  The procedure constructs a controlled mock ensemble by generating non-negative spectra, forwarding them through the Euclidean kernel, checking clean-correlator admissibility diagnostics, and adding controlled noise.  A reconstruction method enters through the uncertainty law, sample ensemble, or interval construction it reports for specified summaries.  The same empirical coverage, simulation-based calibration (SBC), physical diagnostics, and stress tests are then applied to these reported uncertainty claims \cite{Cook:2006uq,Talts:2020sbc,Stuart:2010bayesinv}.

Existing comparisons of spectral reconstructions often examine central spectra, posterior or bootstrap spreads, resolution widths, or ansatz variations.  These diagnostics are useful, but they do not by themselves determine whether a nominal interval for a specified functional \(T[\rho]\) has the advertised repeated-experiment coverage under a Euclidean-admissible data-generating process.  The proposed calibration test is therefore complementary to, rather than a replacement for, channel-specific spectral reconstruction.

This distinction is especially important for Backus--Gilbert-type constructions.  Modern BG and BG-related Gaussian-process formulations naturally define resolution-smeared spectral densities or spectral densities under algorithm-dependent smearing kernels, and the smearing target should be specified before interpreting the resulting uncertainty \cite{Hansen:2019idp,Bailas:2020qmv,DelDebbio:2025bg}.  The local peak-height tests below are therefore deliberately stringent diagnostics of pointwise spectral-feature claims.  They are not statements that a BG method should be judged primarily by an unsmeared peak height.

We first use a controlled Euclidean-admissible benchmark to expose summary-dependent calibration with four specified uncertainty reports: a score-based sampler, MEM-based and BR-based local uncertainty reports, and a BG-style linear-Gaussian report.  The purpose is not to rank the corresponding method families, but to run a closure test of the adequacy procedure and to show that heterogeneous uncertainty reports can be tested by the same target-wise criterion.  We then apply the same logic to a real finite-temperature shear correlator.  A \(\chi^2\) scan first identifies fixed BG-style reconstruction settings that satisfy Euclidean compatibility.  Inside this compatible family, mock calibration is used to select and interpret uncertainty reports for \(\Wlow\), and the same scan is used to test whether the corresponding report can support pointwise peak-height claims.  The shear example illustrates the central distinction: Euclidean compatibility is necessary, but it is neither an uncertainty-calibration criterion nor a functional-adequacy criterion.

The mock ensemble is not a QCD prior; all calibration statements are conditional on the chosen spectral family, kernel, covariance model, target functional, grid, and reconstruction setting.  It is a controlled mock ensemble, parallel to lattice data in the limited but essential sense that spectra are non-negative, clean correlators are images of those spectra under the chosen kernel, and the correlators satisfy explicit Euclidean structural diagnostics.  At zero temperature we use complete-monotonicity-type and positive-semidefinite diagnostics motivated by the Laplace representation of a non-negative measure \cite{Osterwalder:1973dx,Osterwalder:1974tc,Hausdorff:1921momentI,Widder:1941laplace}.  At finite temperature we use bosonic reflection symmetry and thermal positivity diagnostics appropriate to the thermal kernel \cite{Martin:1959kms,Lowdon:2022thermal}.

\section{Protocol}
\label{sec:protocol}

For each mock spectrum, the known value \(T[\rho_{\rm true}]\) and the reported interval \(I_\alpha(T)\) define a direct coverage test.  The role of the mock ensemble is to make this repeated-experiment statement well defined under a specified kernel, covariance model, spectral family, and target functional.  The controlled mock ensemble is
\begin{equation}
  \rhovec_{\rm true}\sim p_{\rm mock}(\rhovec),\quad
  \Gvec_{\rm obs}=\Kmat\rhovec_{\rm true}+\bm\epsilon,
  \quad
  \bm\epsilon\sim {\cal N}(0,\Sig),
  \label{eq:mock_world}
\end{equation}
where the weighted matrix \(\Kmat\) includes the quadrature weights.  The stored truth, clean correlator, noisy correlator, covariance, and grids define the calibration problem.  Clean correlators define membership in the controlled mock ensemble; noisy correlators are the inputs to reconstruction.  Calibration statements are therefore conditional on \(p_{\rm mock}\), the kernel, the grid, \(\Sig\), the selected summaries, and the reconstruction setting being tested.

To compare heterogeneous methods, we use only the uncertainty object they report.  We call this interface an adapter,
\begin{equation}
  {\cal A}_{\lambda}: (\Gvec_{\rm obs},\tau,\omega,\Sig,{\cal C})
  \longmapsto (U_{\lambda},D_{\lambda}),
  \label{eq:adapter_map}
\end{equation}
where \(\lambda\) denotes method settings and \({\cal C}\) collects shared physical metadata.  The uncertainty report \(U_\lambda\) can be a posterior sample set, a local Gaussian approximation, a linearly propagated Gaussian law, a bootstrap ensemble, or another report that can be mapped to target-summary intervals.  The protocol does not require the internal philosophies of MEM, BR, BG, or a score model to be identical.  It records the uncertainty claim made by each reconstruction prescription and tests that claim.

\begin{figure*}[t]
  \centering
  \includegraphics[width=0.95\textwidth]{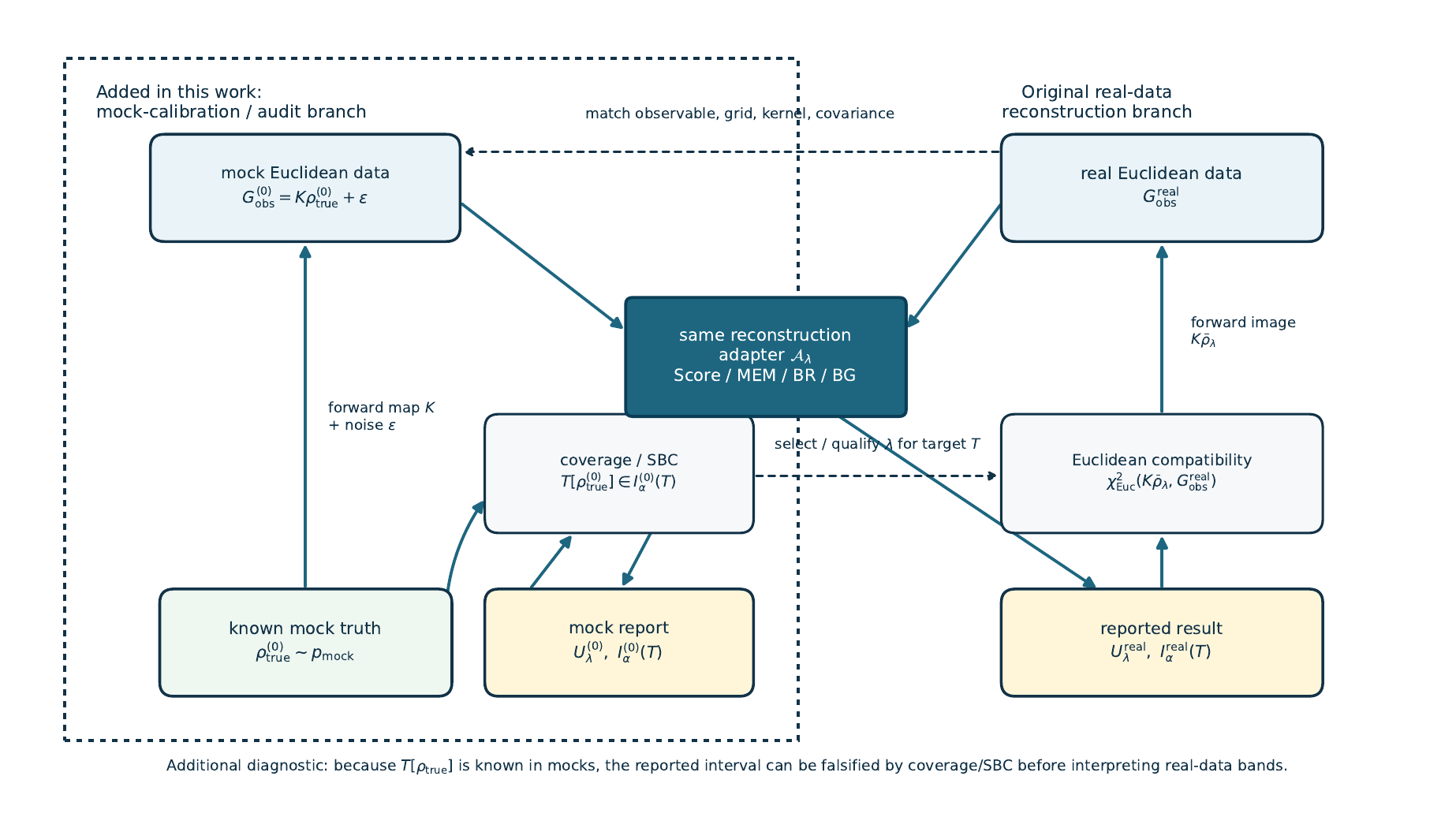}
  \caption{Protocol workflow.  The left dashed box is the added mock-calibration branch: spectra are sampled from the mock generator, forwarded through the Euclidean kernel, reconstructed with the same adapter, and compared against the known truth by coverage, SBC, and diagnostic tests.  The right branch is the original real-data reconstruction path: the measured correlator is reconstructed with the same adapter and checked for Euclidean compatibility.  The real-data uncertainty report is interpreted only after the corresponding mock branch has tested the target-specific interval claim.}
  \label{fig:protocol_workflow}
\end{figure*}

Figure~\ref{fig:protocol_workflow} summarizes the operational role of this added mock-calibration branch relative to a direct real-data reconstruction.

In the main comparisons every adapter is reduced to samples \(\{\rho^{(s)}\}_{s=1}^S\) and to central intervals for
\begin{equation}
  T(\rho)\in\{\wpeak,\rpeak,\Wlow(3)\},
  \quad
  \rpeak=\max_k\rho_k .
  \label{eq:targets}
\end{equation}
The zero-temperature value \(\omega_c=3\) is a predeclared target definition.  Changing \(\omega_c\) changes the target and requires a new calibration test rather than an extrapolation of the old one.  The empirical coverage of a nominal central interval \(I_{\alpha,n}(T)\) is
\begin{equation}
  \widehat C_{\alpha}(T)
  =\frac{1}{N}\sum_{n=1}^N
  {\bf 1}\!\left[T(\rho_{{\rm true},n})\in I_{\alpha,n}(T)\right].
  \label{eq:coverage}
\end{equation}
For sample-level reports we also compute SBC ranks of the true summary among the sampled summary values.  Calibrated sample reports should give approximately uniform ranks \cite{Cook:2006uq,Talts:2020sbc}.  Pointwise spectral bands, when displayed, are interpreted only as pointwise quantile bands, not as simultaneous confidence bands over the full function.

Physical diagnostics are evaluated in addition to coverage.  Central spectra are propagated back to Euclidean time, and we record normalization drift, zero-temperature monotonicity and convexity violations, finite-temperature reflection-symmetry deviations, and small positive-semidefinite matrix margins.  These diagnostics check whether the reported spectra remain close to the admissible Euclidean manifold used to construct the benchmark; they do not replace coverage tests.

For mock-only demonstrations a finite candidate set \(\Lambda_{\rm cand}\) is selected by a held-out calibration rule.  For real data, the selection is explicitly two-step.  The first requirement is the Euclidean compatibility criterion,
\begin{equation}
  \frac{\chi^2_{\rm Euc}(\lambda)}{N_\tau}
  =\frac{1}{N_\tau}
  \left\|\Sig^{-1/2}\bigl[\Kmat\bar\rho_\lambda-\Gvec_{\rm obs}\bigr]\right\|_2^2
  <\chi^2_{\rm cut},
  \label{eq:chi2_gate}
\end{equation}
For normalized transfer observables, \(\Gvec_{\rm obs}\) and \(\Kmat\bar\rho_\lambda\) in Eq.~\eqref{eq:chi2_gate} denote the corresponding preprocessed observable and forward image, including the fixed ultraviolet convention when present.  Alternatively, one can report the full \(\chi^2\)-calibration map over candidates.  This criterion asks only whether the reported central spectrum is compatible with the Euclidean data under the stated error model.  It does not certify an uncertainty interval.

The second layer is target-specific calibration inside the compatible family:
\begin{equation}
\begin{aligned}
  J_T(\lambda)=L_{\rm cov}(T;\lambda)+\eta_{\rm SBC}L_{\rm SBC}(T;\lambda)+P_{\rm phys}(\lambda),\\
  \lambda\in\Lambda_{\chi^2},
\end{aligned}
\label{eq:target_score}
\end{equation}
where \(\Lambda_{\chi^2}\) denotes the Euclidean-compatible candidates.  The first term measures coverage error for the selected target, the second summarizes rank non-uniformity when ranks are available, and the last penalizes clear physical-diagnostic failures.  Equation~\eqref{eq:target_score} is intentionally target dependent.  A candidate can be acceptable for \(\Wlow\) and unacceptable for \(\rpeak\), or vice versa.

On real data the truth is unknown, so empirical coverage cannot be computed directly on the real correlator.  The calibration information must come from an observable-matched mock ensemble.  The real-data protocol is therefore: construct a matched mock ensemble; evaluate the fixed reconstruction settings on that ensemble; require real Euclidean compatibility for the same settings; and, only inside the compatible family, use the mock calibration to choose or qualify the target-summary uncertainty report.  The closure test considered here is defined by four ingredients: a known truth for \(T[\rho]\), a Euclidean-admissible forward map, a stated noise model, and a reported interval from the reconstruction prescription.  The test should be read as a conditional adequacy test.  Passing the test means that the reported interval has not been falsified within the specified mock extension, covariance model, grid, and target definition.  Failing the test means that the tested uncertainty law is not adequate for the chosen spectral functional under that extension, even if the central spectrum is Euclidean-compatible.

\section{Euclidean-admissible benchmark}
\label{sec:mockdata}

The main benchmark is generated in spectral space.  Spectra are mixtures of one to three positive Gaussian or log-Gaussian components, optionally thresholded, with a flexible non-negative high-frequency tail.  A zeroth-moment control
\begin{equation}
  S_0[\rho]=\int_0^\infty \rho(\omega)\,\dd\omega
  \label{eq:s0_mock}
\end{equation}
keeps the global scale near a target value.  In the soft-normalized family used for diversity studies the measured mean is \(S_0=1.00003\) with standard deviation \(0.00495\).  The effective spectral locations and widths still vary broadly across the ensemble.  This construction is a physically constrained mock ensemble with controlled support, scale, and morphology.  As emphasized above, its calibration statements remain conditional on the chosen spectral family, kernel, covariance, and target functional.

The clean correlator is obtained by applying the appropriate kernel before noise is added.  For zero temperature we use a Laplace kernel.  Clean samples are checked by discrete complete monotonicity and Hankel positive-semidefinite diagnostics.  For the finite-temperature branch the thermal bosonic kernel is used, complete monotonicity is not imposed, and the clean gate instead checks midpoint reflection \(G(\tau)=G(\beta-\tau)\) together with finite-dimensional positivity diagnostics.  In the reported constructive ensemble the clean gates pass at essentially unit rate for both branches.  The admissibility check therefore confirms that the generator lies in the intended domain; it is not a severe rejection filter that rescues pathological proposals.

The fiducial zero-temperature calibration test uses a Toeplitz covariance
\begin{equation}
  \Sigma_{ij}=\sigma^2\exp(-|i-j|a_\tau/\ell),
  \quad \sigma^2=10^{-5},\quad \ell=0.25,
  \label{eq:toeplitz_cov}
\end{equation}
with \(\Ntau=32\).  The stress campaign varies \(\Ntau\), \(\sigma^2\), \(\ell\), the covariance supplied to inference, the spectral-family descriptors, and sampler-side constraints.  Thus the mock data are not a single curve.  They define a controlled Euclidean inverse problem in which Euclidean admissibility, observation noise, and spectral morphology can be changed separately.

\begin{figure}[t]
  \centering
  \includegraphics[width=\columnwidth]{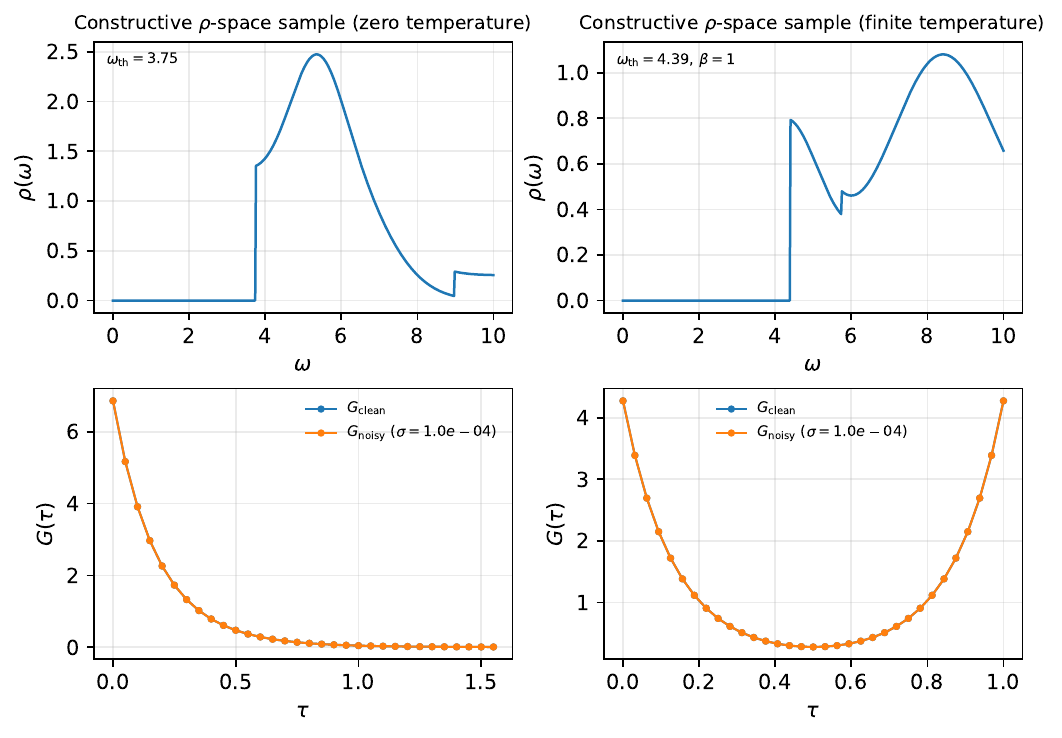}
  \caption{Representative spectra and Euclidean correlators from the constructive benchmark.  The left column illustrates the zero-temperature branch and the right column the finite-temperature branch.  Clean correlators are tested by channel-appropriate admissibility diagnostics before noise is added.}
  \label{fig:mock_universe}
\end{figure}

The generator defines a controlled Euclidean inverse problem with known truth, explicit spectral positivity, channel-dependent clean-correlator admissibility, tunable covariance structure, and recorded morphology labels.  These ingredients provide the setting in which a repeated-experiment statement about an uncertainty band can be tested.  The limitations of the benchmark are also explicit.  It contains positivity, thresholds, a simple normalization control, and generic ultraviolet variation, but it does not encode the detailed physics of every QCD channel.  A transport analysis would ultimately need constraints on \(\rho(\omega)/\omega\) near \(\omega=0\), perturbative or OPE-inspired ultraviolet behavior, known sum rules, contact or subtraction conventions, and the full lattice covariance.

\section{Calibration results}
\label{sec:results}

The purpose of this generic benchmark is methodological: it is a closure test of the calibration procedure and a demonstration that different spectral summaries can have different operational identifiability.  The physics-facing adequacy test is the shear application in Sec.~\ref{sec:realdata}.

The selected representatives are a conditional score sampler, a MEM-based adapter, a BR-based adapter, and a Backus--Gilbert-type adapter.  Throughout this section, the reported coverages quantify the uncertainty laws produced by these specified adapters, not the intrinsic resolving power of the corresponding reconstruction families.  The score model is trained on accepted mock spectra and noisy correlators and sampled with a warm-prior initialization and sampler-side physical projection.  MEM and BR use entropy-regularized optima lifted by local or evidence-weighted Laplace-type uncertainty.  BG uses the Gaussian law induced by linear propagation through the BG reconstruction matrix.  These adapters are described in Appendix~\ref{app:adapters}.  The following comparison concerns these reported uncertainty laws only; it is not a ranking of reconstruction philosophies or of all accepted MEM, BR, and BG implementations.

\begin{table}[t]
\caption{Fiducial 95\% empirical coverage at \((\Ntau,\sigma^2,\ell)=(32,10^{-5},0.25)\).  All representatives use the same held-out cases, summaries, sample budget, and central-interval construction.  Nominal coverage would be 0.95.  These values quantify the reported uncertainty laws generated by the specified adapters, not the intrinsic resolving power of the corresponding reconstruction families.}
\label{tab:fiducial_coverage}
\begin{ruledtabular}
\begin{tabular}{lccc}
Method & \(\wpeak\) & \(\rpeak\) & \(\Wlow\)\\
\hline
Score & 0.824 & 0.633 & 0.312\\
MEM-based & 0.539 & 0.000 & 0.066\\
BG-type & 0.887 & 0.762 & 0.125\\
BR-based & 0.793 & 0.457 & 0.023
\end{tabular}
\end{ruledtabular}
\end{table}

Table~\ref{tab:fiducial_coverage} gives the numerical anchor.  At the fiducial point, the score representative gives 95\% coverages \(0.824\), \(0.633\), and \(0.312\) for \(\wpeak\), \(\rpeak\), and \(\Wlow\), respectively.  BG gives good peak-position coverage, \(0.887\), but only \(0.125\) for \(\Wlow\).  MEM-based and BR-based local uncertainty reports show that smooth entropy-regularized central spectra do not by themselves imply calibrated intervals.  With \(N=256\) cases, the binomial standard error is only a few percent, so the infrared deficits are not Monte Carlo fluctuations.

The rank diagnostics carry the same message for the sample-producing score representative.  Table~\ref{tab:score_sbc_ks} lists the two-sided Kolmogorov--Smirnov statistic of the SBC rank distribution at the fiducial and hard operating points.  The low-frequency ranks are far more nonuniform than the peak-location ranks, even when the 95\% coverage alone is already visibly low.  Thus the infrared problem is not only a tail-interval effect; the full sampled distribution is displaced relative to the mock truth.

Table~\ref{tab:wlow_cutoff_rescan} gives a nearby-cutoff rescan for the low-frequency weight.  The same fixed representatives are evaluated at \(\omega_c=2,3,4\), without retraining or retuning.  The numerical severity changes with the cutoff, but nominal 95\% coverage is not restored in either the fiducial or hard setting.  Across the entries in Table~\ref{tab:wlow_cutoff_rescan}, the largest 95\% coverage is only \(0.453\), and most values are much smaller.  The corresponding rank nonuniformity remains large: the \(D_{\rm KS}\) values range from \(0.349\) to \(0.984\) over the same rescan.  This rules out the interpretation that the \(\Wlow(3)\) undercoverage is an accident of the single predeclared cutoff.

\begin{table}[t]
\caption{Representative SBC rank diagnostics for the selected score-based uncertainty report.  The columns \(D_{\rm KS}\) are two-sided Kolmogorov--Smirnov distances of the rank distribution from uniformity.  The same cases also give the raw 95\% coverages shown in the last three columns.}
\label{tab:score_sbc_ks}
\begin{ruledtabular}
\begin{tabular}{lcccccc}
Setting & \multicolumn{3}{c}{\(D_{\rm KS}\)} & \multicolumn{3}{c}{\(C_{0.95}\)}\\
 & \(\wpeak\) & \(\rpeak\) & \(\Wlow\) & \(\wpeak\) & \(\rpeak\) & \(\Wlow\)\\
\hline
\(P_{\rm mid}\) & 0.149 & 0.280 & 0.652 & 0.824 & 0.633 & 0.312\\
\(P_{\rm hard}\) & 0.219 & 0.340 & 0.672 & 0.555 & 0.441 & 0.207
\end{tabular}
\end{ruledtabular}
\end{table}

\begin{figure*}[t]
  \centering
  \includegraphics[width=0.8\textwidth]{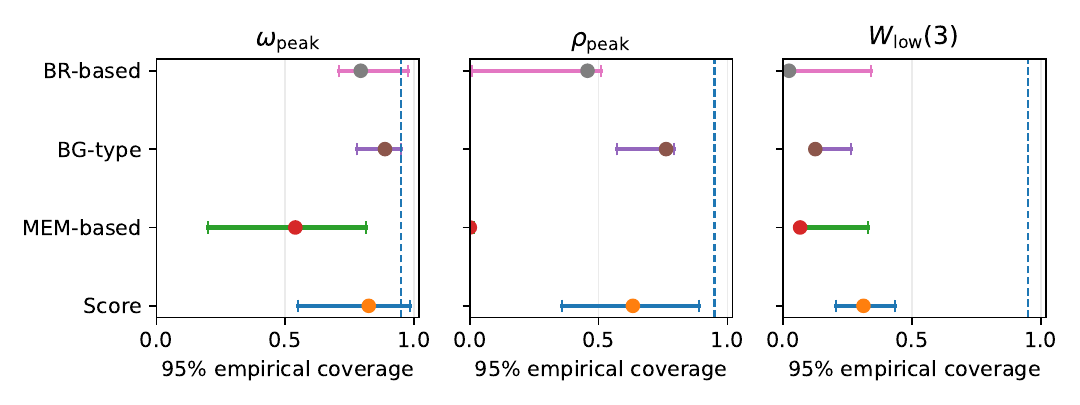}
  \caption{Summary of the primary calibration stress matrix over the \(3\times 3\) grid in \((\Ntau,\sigma^2)\) at fixed Toeplitz correlation length \(\ell=0.25\).  Each horizontal interval shows the range of 95\% empirical coverage over the stress grid, the marker shows the fiducial point \((\Ntau,\sigma^2,\ell)=(32,10^{-5},0.25)\), and the dashed vertical line marks nominal 0.95 coverage.  Peak position is the most calibratable summary, whereas the low-frequency weight remains strongly undercovered across methods.}
  \label{fig:stress_summary}
\end{figure*}

\begin{table*}[t]
\caption{Cutoff-rescan sanity check for the low-frequency weight.  The entries are 95\% empirical coverages for
\(\Wlow(\omega_c)=\int_0^{\omega_c}\rho(\omega)\dd\omega\), evaluated with the same fixed representatives selected for
the original \(\omega_c=3\) audit.  No retraining or retuning is performed.  The two settings are the fiducial
\(P_{\rm mid}=(N_\tau,\sigma^2,\ell)=(32,10^{-5},0.25)\) and the harder
\(P_{\rm hard}=(16,10^{-4},0.25)\).}
\label{tab:wlow_cutoff_rescan}
\begin{ruledtabular}
\begin{tabular}{lcccccc}
Method
& \multicolumn{3}{c}{\(P_{\rm mid}\)}
& \multicolumn{3}{c}{\(P_{\rm hard}\)}\\
& \(\omega_c=2\) & \(\omega_c=3\) & \(\omega_c=4\)
& \(\omega_c=2\) & \(\omega_c=3\) & \(\omega_c=4\)\\
\hline
Score     & 0.125 & 0.312 & 0.422 & 0.129 & 0.207 & 0.359\\
MEM-based & 0.074 & 0.066 & 0.176 & 0.117 & 0.262 & 0.453\\
BG-type   & 0.063 & 0.125 & 0.336 & 0.113 & 0.207 & 0.293\\
BR-based  & 0.000 & 0.023 & 0.000 & 0.016 & 0.059 & 0.078
\end{tabular}
\end{ruledtabular}
\end{table*}

The range summary in Fig.~\ref{fig:stress_summary} supports the same conclusion while leaving the dense cell-by-cell stress matrix to Appendix~\ref{app:full_stress}.  For the score representative, the 95\% coverage of \(\wpeak\) ranges from \(0.551\) to \(0.984\), whereas the corresponding coverage of \(\Wlow\) ranges only from \(0.207\) to \(0.434\).  The specified classical adapters fail differently: the BG-type uncertainty report is closer to nominal for peak localization, while the local MEM and BR uncertainty lifts are often much too narrow for peak height and infrared weight.  The full matrices and the original heat-map visualization are collected in Appendix~\ref{app:full_stress}.

The benchmark reveals a hierarchy of operational functional identifiability.  A peak displacement changes the Euclidean correlator in a relatively coherent way.  The low-frequency integral is much less localized in data space.  It changes under threshold shifts, broadening, redistribution within the infrared region, and compensation against the fixed total normalization.  Positivity and the \(S_0\) projection couple the infrared region to the rest of the spectrum.  Thus a method can report useful uncertainty for a dominant peak position while being overconfident for an infrared proxy under the stated mock ensemble and covariance model.

Covariance and family shifts further show that calibration is conditional.  If data are generated with the fiducial Toeplitz covariance but inference is supplied with \(\Sigma'=a\Sigma\), the score representative loses coverage for both peak and infrared summaries.  Supplying only \(\mathrm{diag}(\Sigma)\) is less damaging for the score model in the tested setting, but it can be more harmful for MEM and BR, whose likelihoods and local uncertainty approximations depend directly on the covariance geometry.  The calibration test therefore turns covariance assumptions into explicit inputs rather than implicit choices.

The out-of-family test is the sharpest stress test.  Training and selection are performed on a family with lower thresholds, fewer peaks, and steeper tails, and the calibration test is moved to spectra with higher thresholds, more peaks, and softer tails.  At the fiducial point the threshold-only shift changes the score coverage for \(\Wlow\) from \((C_{0.68},C_{0.95})=(0.777,0.840)\) in family to \((0.098,0.137)\) out of family.  This identifies threshold support as a leading variable controlling the infrared summary in the present benchmark.  Peak multiplicity alone is not the dominant cause of the observed failure.

Several ablations rule out simpler explanations.  Removing individual sampler-side Euclidean filters changes the selected score representative only mildly once the training spectra already satisfy the clean gates.  Dropping the \(S_0\) control produces visible normalization drift but does not restore infrared calibration.  Increasing an effective posterior temperature widens some intervals, but \(\Wlow\) remains far below nominal coverage and peak localization degrades.  The low-frequency failure is therefore not a plotting artifact or a single tuning mistake.  It is a weakly identifiable direction exposed by the mock calibration test.

All zero-temperature statements about \(\Wlow\) use the predeclared cutoff \(\omega_c=3\) in the working units of the archived runs.  The conclusion is not that every possible infrared functional has the same numerical coverage.  It is that the tested uncertainty reports are not certified for this fixed transport-motivated infrared summary.  The nearby-cutoff rescan in Table~\ref{tab:wlow_cutoff_rescan} treats \(\omega_c=2\) and \(4\) as new summaries rather than extrapolating from \(\omega_c=3\).  The result strengthens the infrared conclusion: while the numerical coverage is cutoff dependent, the tested reports remain far below nominal coverage throughout this local rescan.  This keeps the infrared claim falsifiable and avoids turning one calibrated target into an unstated continuum of targets.  A channel-specific transport analysis should repeat the same audit for the corresponding dimensionless \(\omega/T\) target or Kubo-limit functional.

\section{Real-data shear transfer: \texorpdfstring{\(\Wlow\)}{Wlow} calibration and peak-height failure}
\label{sec:realdata}

We now apply the same calibration test to a real finite-temperature shear correlator.  We use the released shear correlator as a real-data transfer test of functional adequacy.  The target is the reliability of reported spectral-summary intervals under the stated observable matching and covariance assumptions, rather than a determination of \(\eta/s\) or a reproduction of the full source-analysis transport extraction.  The test has two requirements.  First, the reported central spectra must be Euclidean-compatible with the released correlator.  Second, within that compatible family, the uncertainty budgets for a chosen target summary are selected or rejected using calibration measured on an observable-matched mock ensemble.  This section therefore asks which physical claims are supported by the tested BG-style uncertainty law once Euclidean compatibility has already been imposed.

The observable-matched benchmark follows the normalized shear channel of Ref.~\cite{Altenkort:2023eav}.  Synthetic finite-temperature spectra are propagated with the bosonic kernel, sliced to the retained \(\tau T\) points, and normalized to
\begin{equation}
  y(\tau)=\frac{G(\tau)}{G_{\rm norm}(\tau)} .
  \label{eq:real_y}
\end{equation}
The finite-temperature target tested below is not the zero-temperature \(\Wlow(3)\) of Sec.~\ref{sec:results}.  We define
\begin{equation}
  \Wlow^{\rm sh}(\Omega_c)=\int_0^{\Omega_c}\rho_{\rm sh}^{\rm work}(\Omega)\,\dd\Omega,
  \qquad \Omega=\omega/T,
  \qquad \Omega_c=3,
  \label{eq:real_wlow_target}
\end{equation}
where \(\rho_{\rm sh}^{\rm work}\) is the spectrum in the normalized-observable, fixed-ultraviolet convention of the transfer setup.  This infrared weight is used to test an uncertainty report in the working normalized convention; it is not a Kubo-limit estimator of the shear viscosity.

The published pointwise errors are used as a heteroscedastic diagonal covariance for this normalized observable.  This diagonal approximation is a limitation: the \(\chi^2\) values below are standardized residual diagnostics under the working error model, not full goodness-of-fit probabilities based on a complete covariance matrix.

\begin{table*}[t]
\caption{Matching choices for the finite-temperature shear transfer.  The comparison is at the level of the reported uncertainty law for the Euclidean observable, rather than a full reproduction of the source shear-viscosity extraction.}
\label{tab:paper_matching_choices}
\begin{ruledtabular}
\scriptsize
\begin{tabular}{lll}
Ingredient & Source analysis & Present calibration test\\
\hline
Observable & normalized shear correlator & \(y(\tau)=G/G_{\rm norm}\)\\
Time support & retained continuum points & 11 shear points\\
Errors & pointwise errors & diagonal heteroscedastic \(\Sig\)\\
Reconstruction class & resolution cross-check & fixed BG-style linear-Gaussian reports\\
Not matched & transport systematics & full covariance, sum rules, continuum/flow-time systematics, full IR/UV ansatz scans
\end{tabular}
\end{ruledtabular}
\end{table*}

\begin{figure}[t]
  \centering
  \includegraphics[width=0.85\columnwidth]{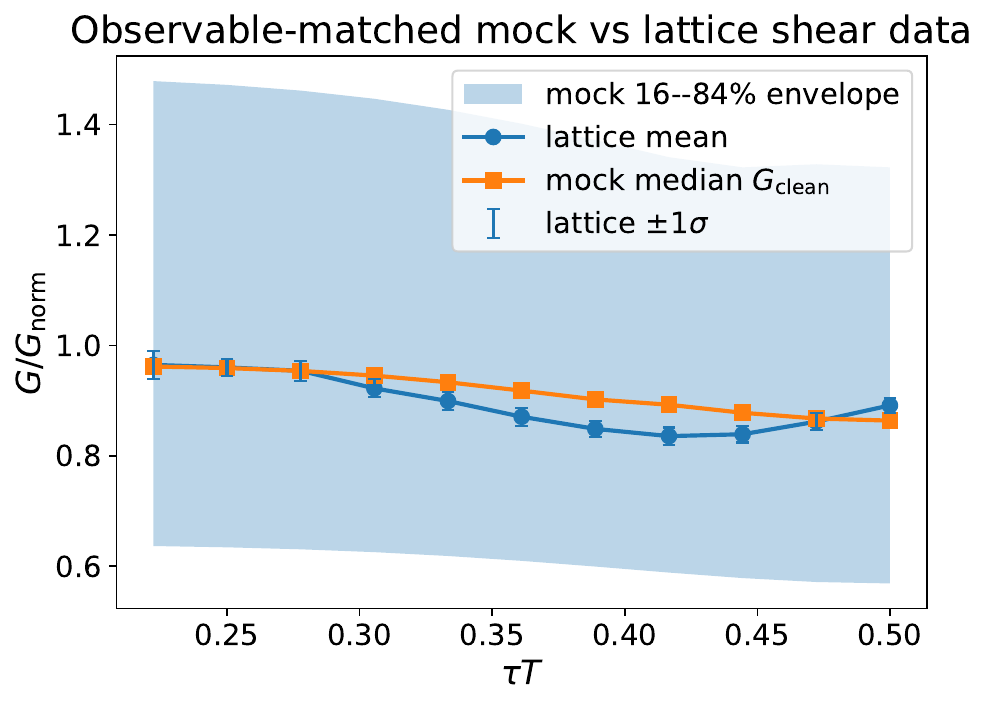}
  \caption{Observable-matched finite-temperature shear benchmark.  The released normalized shear correlator is compared with the mock median and 16--84\% envelope on the retained \(\tau T\) support.  This checks scale and support for the transfer environment; the calibration statement remains conditional on the observable-matched mock ensemble.}
  \label{fig:shear_stage3_vs_paper}
\end{figure}

Figure~\ref{fig:shear_stage3_vs_paper} shows that the real shear observable lies on the same scale as the observable-matched mock benchmark.  As a separate feasibility check, a direct non-negative Euclidean fit under the same observable preprocessing, diagonal errors, finite-temperature kernel, working frequency grid, and fixed ultraviolet convention gives
\begin{equation}
  \chi^2_{\rm data}/N_\tau=0.855,
  \qquad \max_i |z_i|=1.776,
  \label{eq:shear_feasibility}
\end{equation}
for the 11 retained shear points.  The retained real correlator is therefore not outside the working forward setup.  The remaining question is whether the reported uncertainty interval is calibrated for the chosen spectral summary.

We evaluate 8001 fixed BG-style reconstruction settings and compare their Euclidean residuals with their mock-calibration diagnostics.  The smallest Euclidean mismatch is
\begin{equation}
  \min_\lambda \chi^2_{\rm Euc}(\lambda)/N_\tau=1.267,
  \label{eq:real_best_chi2}
\end{equation}
for reconstruction setting 07942.  The scan contains 119 reconstruction settings with \(\chi^2/N_\tau<1.5\), 154 with \(\chi^2/N_\tau<2\), and 160 with \(\chi^2/N_\tau<4\).  This establishes that the real shear correlator admits a nontrivial Euclidean-compatible BG-style family within the scanned set of reconstruction settings.

\begin{figure*}[t]
  \centering
  \includegraphics[width=0.9\textwidth]{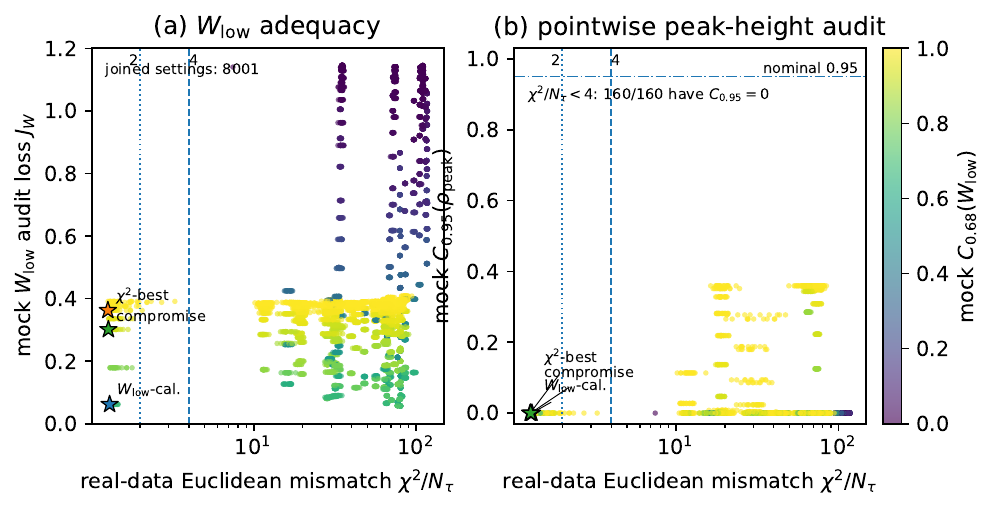}
  \caption{Euclidean compatibility versus target-wise calibration for the 8001 fixed BG-style shear reconstruction settings.  Left: the \(\Wlow\)-calibration loss \(J_W\) in Eq.~\eqref{eq:real_jw_main}, colored by the mock 68\% coverage for \(\Wlow\).  Right: the mock 95\% coverage for the pointwise peak-height target \(\rho_{\rm peak}\), colored by the same \(\Wlow\) coverage.  The dashed and dotted vertical lines indicate \(\chi^2/N_\tau=4\) and 2, respectively, the horizontal reference line in the right panel marks nominal 0.95 coverage, and stars mark the three representative reports in Table~\ref{tab:real_shear_configs}.  In the Euclidean-compatible region, the scan contains calibrated or near-calibrated \(\Wlow\) representatives but no certified pointwise peak-height intervals for the tested BG-style uncertainty law.  Since the \(\chi^2/N_\tau<2\) and \(\chi^2/N_\tau<1.5\) subsets are contained in the displayed \(\chi^2/N_\tau<4\) region, the peak-height failure persists under stricter compatibility cuts.}
  \label{fig:real_shear_chi2_calibration_map}
\end{figure*}

For the map below we use
\begin{equation}
  J_W=\max_{\alpha\in\{0.68,0.95\}}\left|\widehat C_\alpha(\Wlow)-\alpha\right|
  +0.2D_{\rm KS}(\Wlow),
  \label{eq:real_jw_main}
\end{equation}
with no additional physical-diagnostic penalty after invalid settings have been removed.  Thus Fig.~\ref{fig:real_shear_chi2_calibration_map} displays the two validation requirements directly: real-data Euclidean compatibility on the horizontal axis and matched-mock target calibration on the vertical axis.

Figure~\ref{fig:real_shear_chi2_calibration_map} separates two logically distinct requirements: agreement with the Euclidean correlator and calibration of the target-summary interval.  The compatible reconstruction settings do not have identical mock-calibration behavior for \(\Wlow\).  Some reports fit the real Euclidean correlator very well but are overconservative for \(\Wlow\); others have similar real-data \(\chi^2\) but a smaller target-specific calibration loss.  A pure \(\chi^2\) choice would therefore answer only whether the central spectrum can explain the Euclidean data.  It would not answer whether the \(\Wlow\) interval has the intended repeated-experiment interpretation in the matched mock ensemble.

\begin{table*}[t]
\caption{Representative BG-style shear reports inside the low-\(\chi^2\) family.  The real-data \(\Wlow\) columns are computed from the fixed report applied to the released shear correlator.  The \(\Wlow\) diagnostics \(C_{0.68}\), \(C_{0.95}\), and \(D_{\rm KS}\) come from the observable-matched mock calibration.  The final two columns show the peak-height target conflict: the same report family is not certified for \(\rpeak\).}
\label{tab:real_shear_configs}
\begin{ruledtabular}
\scriptsize
\begin{tabular}{lccccccccc}
Role & ID & \(\chi^2/N_\tau\) & \(\Wlow\) median & 68\% width & \(C_{0.68}^{W}\) & \(C_{0.95}^{W}\) & \(D_{\rm KS}^{W}\) & \(C_{0.68}^{\rho_p}\) & \(C_{0.95}^{\rho_p}\)\\
\hline
\(\Wlow\)-calibration & 07802 & 1.298 & 6.295 & 3.114 & 0.668 & 0.961 & 0.250 & 0.000 & 0.000\\
\(\chi^2\)-best & 07942 & 1.267 & 6.345 & 7.155 & 0.980 & 1.000 & 0.310 & 0.000 & 0.000\\
compromise & 07812 & 1.276 & 6.474 & 5.141 & 0.926 & 1.000 & 0.278 & 0.000 & 0.000
\end{tabular}
\end{ruledtabular}
\end{table*}

\begin{table}[t]
\caption{Peak-height failure under progressively stricter Euclidean-compatibility cuts.  For each cut, every retained BG-style report has zero mock coverage for the pointwise peak-height interval.  The calibration size is 256 mock correlators.}
\label{tab:peakheight_cuts}
\begin{ruledtabular}
\begin{tabular}{cccc}
Cut on \(\chi^2/N_\tau\) & Reports & \(C_{0.68}^{\rho_p}\) range & \(C_{0.95}^{\rho_p}\) range\\
\hline
\(<1.5\) & 119 & 0--0 & 0--0\\
\(<2\) & 154 & 0--0 & 0--0\\
\(<4\) & 160 & 0--0 & 0--0
\end{tabular}
\end{ruledtabular}
\end{table}

\begin{figure*}[t]
  \centering
  \includegraphics[width=0.8\textwidth]{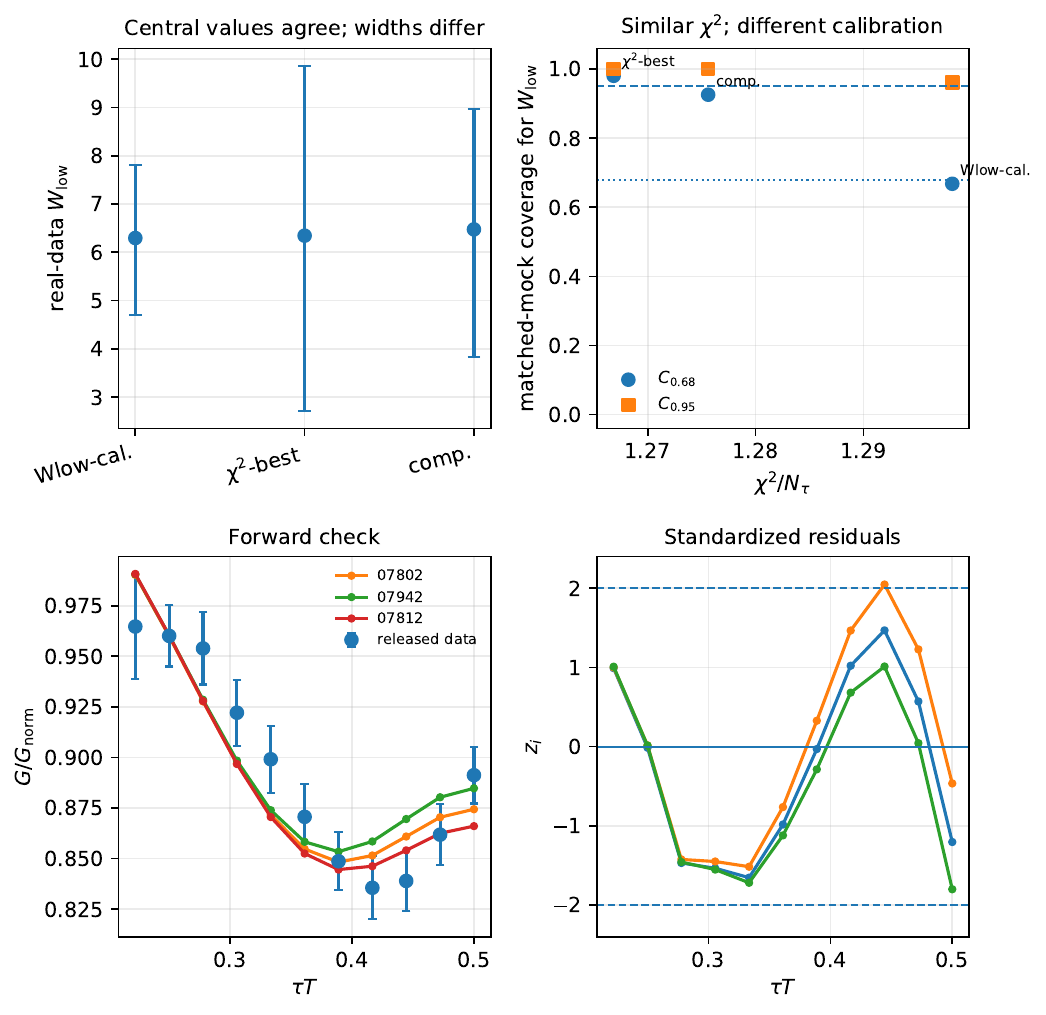}
  \caption{Supporting checks for the three low-\(\chi^2\) shear reports.  The upper-left panel shows real-data \(\Wlow\) medians and 68\% intervals, and the upper-right panel shows the matched-mock coverages for the same reports.  The central values are similar, while the widths differ by more than a factor of two; the \(\chi^2\)-best report is Euclidean-compatible but overcovers \(\Wlow\), whereas the \(\Wlow\)-calibration representative is closer to nominal within the finite scanned grid.  The lower panels give the per-time Euclidean forward check for the same reports: released normalized shear data and sample-mean forward images on the left, and standardized residuals \(z_i=(y_{{\rm model},i}-y_{{\rm data},i})/\sigma_i\) on the right.  Under this sample-mean central-curve convention the recomputed values are \(\chi^2/N_\tau=1.289\), 1.472, and 1.326 for 07802, 07942, and 07812, respectively, with \(\max_i|z_i|\leq 2.05\).}
  \label{fig:real_shear_supporting_checks}
\end{figure*}

Table~\ref{tab:real_shear_configs} and Fig.~\ref{fig:real_shear_supporting_checks} give the uncertainty-budget outcome and the per-time Euclidean residual check.  The three representatives have nearly the same \(\Wlow\) central value, around 6.3--6.5 in the working units, and all have \(\chi^2/N_\tau\simeq 1.3\) under the archived scalar compatibility criterion.  Their 68\% widths, however, range from 3.11 to 7.15.  The selected reports therefore give similar real-data \(\Wlow\) medians but substantially different interval widths, showing that the uncertainty claim is not determined by the Euclidean central fit alone.  The \(\chi^2\)-best setting overcovers the target in the matched mock calibration.  The \(\Wlow\)-calibrated representative has \(C_{0.68}=0.668\) and \(C_{0.95}=0.961\), close to nominal at the finite calibration size, while remaining Euclidean-compatible on the real correlator.  The forward residuals in Fig.~\ref{fig:real_shear_supporting_checks} show that this compatibility is not produced by hiding a single large time-slice discrepancy: for the displayed sample-mean central spectra, all retained residuals are at about the two-standard-deviation level or smaller.  Among reconstruction settings with \(\chi^2/N_\tau<4\), the selected report is one representative of the minimum-\(J_W\) equivalence class at the finite mock-sample resolution.  We therefore use it as the calibration-selected \(\Wlow\) representative within the scanned grid and stated mock calibration, not as a claim of globally optimal uncertainty quantification.

The peak-height diagnostic gives the sharper adequacy test for local spectral structure.  In the low-\(\chi^2\) representatives of Table~\ref{tab:real_shear_configs}, the peak-height coverages are \(C^{\rho_p}_{0.68}=C^{\rho_p}_{0.95}=0\).  The full scan in Fig.~\ref{fig:real_shear_chi2_calibration_map} and Table~\ref{tab:peakheight_cuts} shows that this is not a consequence of choosing three unlucky examples: throughout the Euclidean-compatible region \(\chi^2/N_\tau<4\), all 160 BG-style reports have \(C^{\rho_p}_{0.95}=0\) in the observable-matched mock calibration, and the same statement holds for the stricter \(\chi^2/N_\tau<2\) and \(\chi^2/N_\tau<1.5\) cuts.  With 256 mock correlators, zero 95\% coverage is far outside a finite-binomial fluctuation around nominal coverage, for which \(\sqrt{0.95(1-0.95)/256}=0.014\).  Thus the tested BG-style linear-Gaussian uncertainty law can describe the released shear correlator and can identify a \(\Wlow\)-calibrated representative within the scanned grid, but it does not support calibrated claims about the pointwise peak height.  This result is not a theorem about all Backus--Gilbert constructions.  It is a conditional rejection of the present uncertainty report for a local spectral functional under the stated observable-matched mock extension.

This is also consistent with the resolution-based nature of BG-type methods.  The pointwise peak height is a stringent local diagnostic; a resolution-matched smeared peak amplitude would be a less local target and should be audited separately.  The present result therefore constrains local peak-height claims made from this uncertainty law, rather than the use of BG-type methods for smeared spectral quantities.

The existence of a \(\Wlow\)-calibrated low-\(\chi^2\) representative in the shear example does not contradict the generic zero-temperature undercoverage result of Sec.~\ref{sec:results}.  The two calibration tests use different mock ensembles, target conventions, and scanned reconstruction settings.  This comparison shows that \(\Wlow\) calibration is conditional and must be tested in the target environment.

The shear example demonstrates that a reconstruction can be Euclidean-compatible while its interval for a chosen spectral summary is miscalibrated or uncertified.  The shear correlator is feasible under the working forward setup, a sizable BG-style family satisfies the \(\chi^2\) compatibility criterion, and the mock-calibration test identifies a \(\Wlow\)-calibrated representative within the scanned grid that a pure \(\chi^2\) rule would not select.  At the same time, the same low-\(\chi^2\) family fails the pointwise peak-height audit.  Therefore \(\chi^2\) is a consistency requirement on the forward correlator, not a calibration criterion for spectral uncertainty or local spectral structure.  The transfer test remains conditional on the diagonal covariance, the observable-matched mock ensemble, and the absence of full transport systematics such as known sum rules, continuum and flow-time systematics, full IR/UV ansatz scans, and a Kubo-limit target.

\section{Conclusion}
\label{sec:conclusion}

We have formulated a conditional model-adequacy test for spectral uncertainty claims in lattice QCD.  The central question is whether a nominal interval has the advertised repeated-experiment meaning for the physical summary being interpreted.  The protocol is the operational realization of this question: it constructs a Euclidean-admissible mock ensemble with known truth, maps heterogeneous reconstruction outputs to reported uncertainty laws, and tests empirical coverage, SBC ranks, physical diagnostics, and stress stability.  The test is intentionally asymmetric: passing does not prove QCD truth, but failing invalidates the tested uncertainty claim for the chosen functional within the stated mock extension.

The generic benchmark shows that calibration follows a hierarchy of operational functional identifiability.  In the tested mock ensemble, peak locations are substantially easier to calibrate than peak heights or the low-frequency weight \(\Wlow(3)\).  This hierarchy has a natural inverse-problem interpretation: coherent peak shifts are more visible in Euclidean time than local amplitudes or infrared redistributions that can be compensated elsewhere in the spectrum.  The result should be read conditionally, not as a universal ranking of reconstruction families.

The shear transfer gives the physics-facing adequacy test.  The released shear correlator admits a nontrivial low-\(\chi^2\) BG-style family, so the tested reports are not rejected by Euclidean compatibility alone.  Within the scanned grid and stated observable-matched mock extension, the \(\Wlow\) uncertainty budget is target dependent and a calibrated representative can be identified by mock calibration.  However, the same Euclidean-compatible reports do not certify the pointwise peak-height summary.  This identifies a limitation of the tested BG-style uncertainty law for local spectral-feature claims: good Euclidean compatibility does not imply that all spectral functionals are reliably resolved.

The test is portable because it acts on the reported uncertainty claim itself.  It requires only that the reconstruction output can be mapped to intervals or samples for \(T[\rho]\), independent of the internal reconstruction philosophy.  Future applications should replace the generic mock families by channel-specific benchmarks, include full covariance matrices and known sum rules, audit Kubo-limit and resolution-matched smeared targets directly, and extend the selection to multi-correlator and continuum-limit settings.  Spectral uncertainty bands should be validated for the physical summaries they are used to support.

\section*{Data and code availability}

A reproducibility package supporting this manuscript has been archived on Zenodo under DOI \href{https://doi.org/10.5281/zenodo.20606367}{10.5281/zenodo.20606367}.  It contains the dataset-generation scripts, adapter definitions, coverage-evaluation entry points, reconstruction-grid metadata, plotting scripts, and the joined shear \(\chi^2\)-calibration table used in Fig.~\ref{fig:real_shear_chi2_calibration_map} and Table~\ref{tab:peakheight_cuts}.  The real-shear calibration test uses the released flow-extrapolated shear correlator of Ref.~\cite{Altenkort:2023eav}.  The ancillary package includes the archived BG-style \(\chi^2\) scan, mock-calibration summaries, peak-height coverage columns, selected real-data \(\Wlow\) summaries, and per-time residual diagnostics needed to reproduce Figs.~\ref{fig:real_shear_chi2_calibration_map} and \ref{fig:real_shear_supporting_checks}.

\begin{acknowledgments}
The author thanks members of the Center for High Energy Physics, Peking University, for helpful discussions. The numerical experiments were carried out using local computing resources. Generative AI tools were used solely to assist with language editing for grammar, wording, and readability. All scientific ideas, data generation, calculations, figures, references, and final interpretations were produced and verified by the author.
\end{acknowledgments}

\appendix

\section{Kernels and admissibility diagnostics}
\label{app:kernels_diagnostics}

This appendix records the Euclidean kernels and the clean-correlator diagnostics used to define the benchmark.  These checks are practical finite-dimensional diagnostics, not a complete axiomatization of QCD correlators.

At zero temperature we use
\begin{equation}
  G(\tau)=\int_0^\infty \dd\omega\, e^{-\omega\tau}\rho(\omega),
  \label{eq:zeroT_kernel_app}
\end{equation}
with \(\rho(\omega)\ge 0\) in the benchmark channels.  On the working grid,
\begin{equation}
  G_i=\sum_{k=1}^{K} w_k e^{-\omega_k\tau_i}\rho_k
  \equiv \sum_k \widetilde K_{ik}\rho_k .
  \label{eq:zeroT_discrete_app}
\end{equation}
At finite temperature, for the bosonic transfer study,
\begin{equation}
  K_\beta(\tau,\omega)=
  \frac{\cosh[\omega(\tau-\beta/2)]}{\sinh(\beta\omega/2)},
  \label{eq:finiteT_kernel_app}
\end{equation}
which implies \(G(\tau)=G(\beta-\tau)\).

For the zero-temperature Laplace kernel, a non-negative measure implies complete monotonicity,
\begin{equation}
  (-1)^n\frac{\dd^nG(\tau)}{\dd\tau^n}\ge 0.
  \label{eq:cm_cont_app}
\end{equation}
On the discrete grid we use forward differences and require
\begin{equation}
  (-1)^n\Delta^nG_i\ge -\epsilon_n^{\rm CM},
  \qquad n=0,\ldots,n_{\rm max},
  \label{eq:cm_discrete_app}
\end{equation}
with
\begin{equation}
  \epsilon_n^{\rm CM}=a_{\rm tol}+r_{\rm tol}\max_i|\Delta^nG_i|.
  \label{eq:cm_tol_app}
\end{equation}
Unless otherwise stated \(n_{\rm max}=\min(10,\Ntau-1)\).  Complete monotonicity is applied to clean zero-temperature correlators and is not imposed in the thermal branch.

We also monitor finite-dimensional positive-semidefinite matrices.  A Hankel matrix of depth \(m\) is
\begin{equation}
  H_{ab}=G_{a+b},\qquad a,b=0,\ldots,m-1,
  \label{eq:hankel_app}
\end{equation}
and a Toeplitz matrix is
\begin{equation}
  T_{ab}=g_{|a-b|}.
  \label{eq:toeplitz_app}
\end{equation}
The matrices are symmetrized before diagonalization, and we record the smallest eigenvalue.  These PSD tests are diagnostics of the finite benchmark, not necessary and sufficient conditions for the full physical correlator cone.

The default clean gates are
\begin{equation}
  {\cal G}_{0}[G]=
  {\bf 1}[{\rm CM}(G)\wedge {\rm HankelPSD}(G)]
  \label{eq:gate_zero_app}
\end{equation}
for zero temperature and
\begin{equation}
\begin{split}
  {\cal G}_{\beta}[G]
  &=
  {\bf 1}\bigl[
  {\rm Palindrome}(G)\wedge{} \\
  &\qquad
  ({\rm HankelPSD}(G)\vee {\rm ToeplitzPSD}(G))
  \bigr].
\end{split}
\label{eq:gate_beta_app}
\end{equation}
for finite temperature.  The gate is applied primarily to \(G_{\rm clean}\).  Direct noisy-sequence tests are useful stress diagnostics but are not used to define the physical mock universe.

A projection-based correlator control route is also implemented.  It alternates structured PSD or complete-monotonicity projections with a non-negative least-squares realizability check,
\begin{equation}
  \widehat\rho=\arg\min_{\rho_k\ge 0}\|G-\widetilde K\rho\|_2^2,
  \qquad
  r_{\rm rel}=\frac{\|G-\widetilde K\widehat\rho\|_2}{\|G\|_2+10^{-16}}.
  \label{eq:nnls_app}
\end{equation}
This route is used as a control study to distinguish Euclidean structural repair from realizability by a non-negative spectrum on the chosen grid.

\section{Mock ensemble construction and validation}
\label{app:mock_validation}

The constructive benchmark uses mixtures of Gaussian and log-Gaussian components.  A Gaussian component has
\begin{equation}
  \rho_{\rm G}(\omega)=A\exp\left[-\frac{(\omega-\mu)^2}{2\sigma^2}\right],
  \label{eq:rho_gauss_app}
\end{equation}
with
\begin{equation}
\begin{aligned}
  \mu    &\sim {\rm Unif}(0.2\omega_{\max},0.95\omega_{\max}),\\
  \sigma &\sim {\rm Unif}(0.03\omega_{\max},0.15\omega_{\max}),\\
  A      &\sim {\rm Unif}(0.3,2.0).
\end{aligned}
\label{eq:gauss_ranges_app}
\end{equation}
A log-Gaussian component has
\begin{equation}
  \rho_{\rm logG}(\omega)=
  A\exp\left[-\frac{(\log\omega-\mu_{\log})^2}{2\sigma_{\log}^2}\right]\Theta(\omega),
  \label{eq:rho_loggauss_app}
\end{equation}
with \(\mu_{\log}=\log u\), \(u\sim {\rm Unif}(0.1\omega_{\max},0.8\omega_{\max})\), \(\sigma_{\log}\sim {\rm Unif}(0.2,0.6)\), and \(A\sim {\rm Unif}(0.2,1.5)\).  The number of components is one, two, or three.  A threshold \(\omega_{\rm th}\sim {\rm Unif}(0,0.6\omega_{\max})\) masks the peak sector below threshold.

An optional non-negative tail,
\begin{equation}
  \rho_{\rm tail}(\omega)=c_{\rm tail}\left(\frac{\omega}{\omega_{\rm ref}}\right)^p
  \Theta(\omega-\omega_{\rm tail}),
  \label{eq:tail_app}
\end{equation}
uses \(c_{\rm tail}\sim {\rm Unif}(0,0.5)\), \(p\sim {\rm Unif}(-2,2)\), \(\omega_{\rm tail}\sim {\rm Unif}(0.5\omega_{\max},0.9\omega_{\max})\), and \(\omega_{\rm ref}=\omega_{\max}\).  This tail is only a flexible ultraviolet tail component.

Hard normalization rescales the spectrum to a target \(S_0^\star\).  The soft-alignment mode multiplies this rescaling by
\begin{equation}
  u\sim {\rm LogNormal}\left(-\frac{\sigma_u^2}{2},\sigma_u\right),
  \qquad \sigma_u=0.005,
  \label{eq:soft_s0_app}
\end{equation}
so that \(\E[u]=1\).  In the 2000-sample soft-alignment report the mean absolute residual from \(S_0^\star=1\) is \(3.9589\times10^{-3}\), the median is \(3.2777\times10^{-3}\), and the 95th percentile is \(9.6817\times10^{-3}\).

\begin{table}[t]
\caption{Validation statistics for the constructive benchmark.  Gate rates refer to clean correlators.  Normalization statistics are absolute residuals from \(S_0^\star=1\).}
\label{tab:mock_validation_app}
\scriptsize
\begin{ruledtabular}
\begin{tabular}{lcc}
Diagnostic & Zero temperature & Finite temperature\\
\hline
Main clean-gate rate & \(1.00000\) & \(1.00000\)\\
Filtered accepted set & \(64/64\) & \(64/64\)\\
Mean \(|S_0-S_0^\star|\) & \(3.9589\times10^{-3}\) & \(3.9589\times10^{-3}\)\\
Median \(|S_0-S_0^\star|\) & \(3.2777\times10^{-3}\) & \(3.2777\times10^{-3}\)\\
95th percentile \(|S_0-S_0^\star|\) & \(9.6817\times10^{-3}\) & \(9.6817\times10^{-3}\)
\end{tabular}
\end{ruledtabular}
\end{table}

A 5000-sample diversity report gives effective spectral mean \(6.62997\) with 5th-to-95th percentile range \(4.44028\) to \(8.34990\), effective width mean \(1.64439\) with range \(0.88787\) to \(2.50641\), and support-width mean \(6.84110\) with range \(4.28571\) to \(9.54990\).  The family therefore spans support locations, widths, and peak configurations while retaining a controlled global scale.

The projection-based correlator control route has main and realizability rates \(0.99400\) for the zero-temperature control and \(1.00000\) for the finite-temperature control in 2000-sample reports.  Accepted zero-temperature samples have maximum relative NNLS residual \(1.79696\times10^{-9}\), and finite-temperature samples have maximum residual \(9.70482\times10^{-9}\).  This route is viable but more delicate than the constructive route, so the latter is used for the main audit.

\section{Reconstruction adapters}
\label{app:adapters}

The adapters below define the uncertainty reports tested in this work.  They expose each method's reported uncertainty law in a common sample-level form without making the methods philosophically identical.  The calibration results should therefore be interpreted as statements about these reported uncertainty laws, not as general theorems about MEM, BR, BG, or score-based reconstruction.

The conditional score representative is trained in scaled variables with a variance-exploding denoising score-matching objective \cite{Hyvarinen:2005score,Vincent:2011dsm,Song:2021sde}.  For noise level \(\sigma\) and \(\xi\sim{\cal N}(0,I)\),
\begin{equation}
  \widetilde\rho_\sigma=\widetilde\rho+\sigma\xi,
  \qquad
  s^\star=-\frac{\widetilde\rho_\sigma-\widetilde\rho}{\sigma^2},
  \label{eq:dsm_target_appC}
\end{equation}
and the network minimizes a weighted denoising score loss.  Sampling uses a predictor--corrector discretization on a decreasing VE schedule.  The selected representative uses warm-prior initialization with \((\sigma_{\min},\sigma_{\max},N_{\rm step})=(0.01,1.0,64)\) and shared physical projection.

The MEM adapter maximizes
\begin{equation}
  Q(\rho;\alpha)=\alpha S(\rho,m)-\frac{1}{2}\chi^2(\rho),
  \label{eq:mem_Q_appC}
\end{equation}
with
\begin{equation}
  S(\rho,m)=\sum_k w_k\left[\rho_k-m_k-\rho_k\log\left(\frac{\rho_k}{m_k}\right)\right]
  \label{eq:mem_entropy_appC}
\end{equation}
and the usual Gaussian likelihood term.  Positivity is enforced through \(\rho_k=m_k\exp(y_k)\).  The reported uncertainty is a stabilized Laplace or evidence-weighted local sample law around the selected solution \cite{Tierney:1986laplace}.  This is the tested MEM-based uncertainty law.

The BR adapter uses the Burnier--Rothkopf functional \cite{Burnier:2013nla},
\begin{equation}
  S_{\rm BR}(\rho,m)=\alpha\sum_k w_k
  \left[1-\frac{\rho_k}{m_k}+\log\left(\frac{\rho_k}{m_k}\right)\right],
  \label{eq:br_entropy_appC}
\end{equation}
again in the exponential parameterization.  The selected branch uses integrated-\(\alpha\) optimization and a Laplace-type uncertainty lift.  This is the tested BR-based uncertainty law.

The BG adapter constructs a linear estimator
\begin{equation}
  \widehat\rho=QG_{\rm obs}
  \label{eq:bg_matrix_appC}
\end{equation}
and reports the induced Gaussian law
\begin{equation}
  q_{\rm BG}(\rho|G_{\rm obs})={\cal N}(QG_{\rm obs},Q\Sigma Q^T).
  \label{eq:bg_gaussian_appC}
\end{equation}
Ranks for BG are therefore calibration diagnostics of this reported linear-Gaussian law.

\begin{table*}[t]
\caption{Selected representatives used in the fiducial audit.}
\label{tab:selected_adapters_appC}
\begin{ruledtabular}
\scriptsize
\begin{tabular}{lll}
Method & Selected representative & Audited uncertainty report\\
\hline
Score & prior initialization, \((0.01,1.0,64)\) VE schedule, bi12 checkpoint & conditional score samples with physical projection\\
MEM-based & SVD search, flat-\(S_0\) default, \(\alpha\in[10^{-4},10^{7}]\), 12 nodes, classic-max & stabilized Laplace/evidence-weighted local samples\\
BR-based & integrated-\(\alpha\) optimization on \([10^{-3},10^{3}]\), 16 nodes, full space & BR-MAP plus integrated-\(\alpha\)/Laplace samples\\
BG-type & \(\lambda=0.5\), ridge \(10^{-16}\), \(\omega\)-stride 2, sample scale 1.25 & linear-Gaussian pseudo-posterior
\end{tabular}
\end{ruledtabular}
\end{table*}

Failed optimizer calls, singular covariance factorizations, invalid samples, and non-finite entries are recorded in the adapter metadata.  A reconstruction setting that appears calibrated only after silent case removal is not treated as a valid uncertainty report.

\section{Metrics and selection}
\label{app:metrics_selection}

For each sampled spectrum we compute
\begin{align}
  \wpeak &=\omega_{\arg\max_k\rho_k},\qquad
  \rpeak=\max_k\rho_k,\nonumber\\
  \Wlow(\omega_c)&=\sum_{\omega_k\le \omega_c}w_k\rho_k .
  \label{eq:app_targets}
\end{align}
The main zero-temperature audit uses \(\omega_c=3\).  If several bins share the maximum within numerical tolerance, the first maximum is used consistently for truth and samples.

Central intervals are empirical quantile intervals,
\begin{equation}
 I_\alpha(T)=
 \left[q_{(1-\alpha)/2}\{T(\rho^{(s)})\},
       q_{(1+\alpha)/2}\{T(\rho^{(s)})\}\right].
 \label{eq:app_intervals}
\end{equation}
The empirical coverage is Eq.~\eqref{eq:coverage}, with binomial standard error
\begin{equation}
  {\rm se}[\widehat C_\alpha]\simeq
  \sqrt{\frac{\widehat C_\alpha(1-\widehat C_\alpha)}{N}} .
  \label{eq:app_binom_se}
\end{equation}
For \(N=256\), this standard error is a few percent for the coverages reported here.

SBC ranks are defined by
\begin{equation}
  r_n(T)=\sum_{s=1}^{S}
  {\bf 1}\!\left[T(\rho^{(s)}_n)<T(\rho_{{\rm true},n})\right],
  \label{eq:app_sbc_rank}
\end{equation}
with randomized tie handling when needed.  We summarize non-uniformity with a Kolmogorov--Smirnov distance after mapping ranks to \([0,1]\).  Coverage tests advertised interval frequency; SBC is sensitive to bias and mass allocation within the reported distribution.

For the generic mock demonstration, selected representatives are chosen on held-out cases by a calibration-first score,
\begin{equation}
 J_{\rm global}(\lambda)=L_{\rm cov}(\lambda)+\eta_{\rm SBC}L_{\rm SBC}(\lambda)+P_{\rm phys}(\lambda),
 \label{eq:app_global_score}
\end{equation}
where
\begin{equation}
 L_{\rm cov}(\lambda)=
 \max_{T\in\{\wpeak,\rpeak,\Wlow\}}
 \max_{\alpha\in\{0.68,0.95\}}
 |\widehat C_\alpha(T;\lambda)-\alpha| .
 \label{eq:app_lcov_global}
\end{equation}
The maximum prevents a severe failure for one physical summary from being hidden by better behavior on another.  The fixed selection budgets are 144 MEM candidates, 400 BR candidates, 432 BG candidates, and a fixed score training/sampling grid.

For real-data transfer the rule is two-layer.  The first requirement is the Euclidean compatibility criterion, Eq.~\eqref{eq:chi2_gate}.  In the shear scan we report the full map and quote counts for several thresholds rather than claiming a unique probabilistic cutoff, because only a diagonal pointwise covariance is available.  The second layer is target-specific:
\begin{align}
 J_T(\lambda)&=
 \max_{\alpha\in\{0.68,0.95\}}|\widehat C_\alpha(T;\lambda)-\alpha|\nonumber\\
 &\quad +\eta_{\rm SBC}D_{\rm KS}(T;\lambda)+P_{\rm phys}(\lambda),
 \qquad \lambda\in\Lambda_{\chi^2}.
 \label{eq:app_target_score}
\end{align}
In the displayed shear map we use \(T=\Wlow\) and \(\eta_{\rm SBC}=0.2\) for visualization of the calibration loss.  Invalid configurations are removed before plotting, and no additional physical-diagnostic penalty is applied in that displayed map; equivalently, \(P_{\rm phys}=0\) for Fig.~\ref{fig:real_shear_chi2_calibration_map}.  The precise numerical ranking should not be overinterpreted beyond the finite candidate set; the qualitative requirement is that Euclidean compatibility and target calibration be reported separately.

The real-data target summary has no known truth.  Therefore real-data coverage is not measured on the released correlator.  It is inherited from an observable-matched mock calibration with the same preprocessing, kernel, grid, and diagonal-error convention.  To reduce selection bias, the candidate grid, target definition, mock-calibration summaries, and real-data \(\chi^2\) diagnostics are treated as separate validation components.  The final real-data report is conditional on all of these choices, and no selected configuration is claimed to be globally optimal outside this finite grid or under a different calibration split.

\subsection*{Dataset splits and selection budget}

The mock calibration uses disjoint roles for training, calibration, and final evaluation.  Table~\ref{tab:dataset_roles} records the sizes that enter the reported generic runs.  The score model is the only representative that requires a training split.  Classical adapters are selected on held-out mock cases using the same target summaries and are then fixed before final evaluation.  The final selected-representative numbers in the main text use the same \(N=256\) held-out cases and \(S=128\) samples per case for all adapters.

The real-shear BG-style grid contains 8001 fixed configurations.  Its mock calibration is evaluated with \(N=256\) mock cases and \(S=128\) samples per case for the stored sweep.  The real \(\Wlow\) intervals in Table~\ref{tab:real_shear_configs} use 4096 samples for the three selected configurations.

\begin{table*}[t]
\caption{Dataset roles and evaluation budgets used in the reported audits.  The detailed train/validation/test sizes refer to the mock splits used by the score model; the final selected-representative audit uses the same held-out cases for all adapters within each benchmark setting.}
\label{tab:dataset_roles}
\begin{ruledtabular}
\begin{tabular}{lll}
Stage & Size or budget & Main use\\
\hline
Toy score split & 10000/2000/2000 & score-mechanism diagnosis\\
Fiducial score split & 40000/8000/8000 & score training and validation\\
Adapter selection & 144 MEM; 400 BR; 432 BG; fixed score grid & choose representative adapters\\
Final selected audit & \(N=256,\ S=128\) & coverage, SBC, physical diagnostics\\
Stress audit & \(N=256,\ S=128\) unless stated otherwise & regime and family shifts\\
Real shear BG grid & 8001 candidates; 3 real reports with 4096 samples & two-layer \(\chi^2\)-calibration test
\end{tabular}
\end{ruledtabular}
\end{table*}

\section{Full stress-test tables}
\label{app:full_stress}

\begin{figure*}[t]
  \centering
  \includegraphics[width=0.9\textwidth]{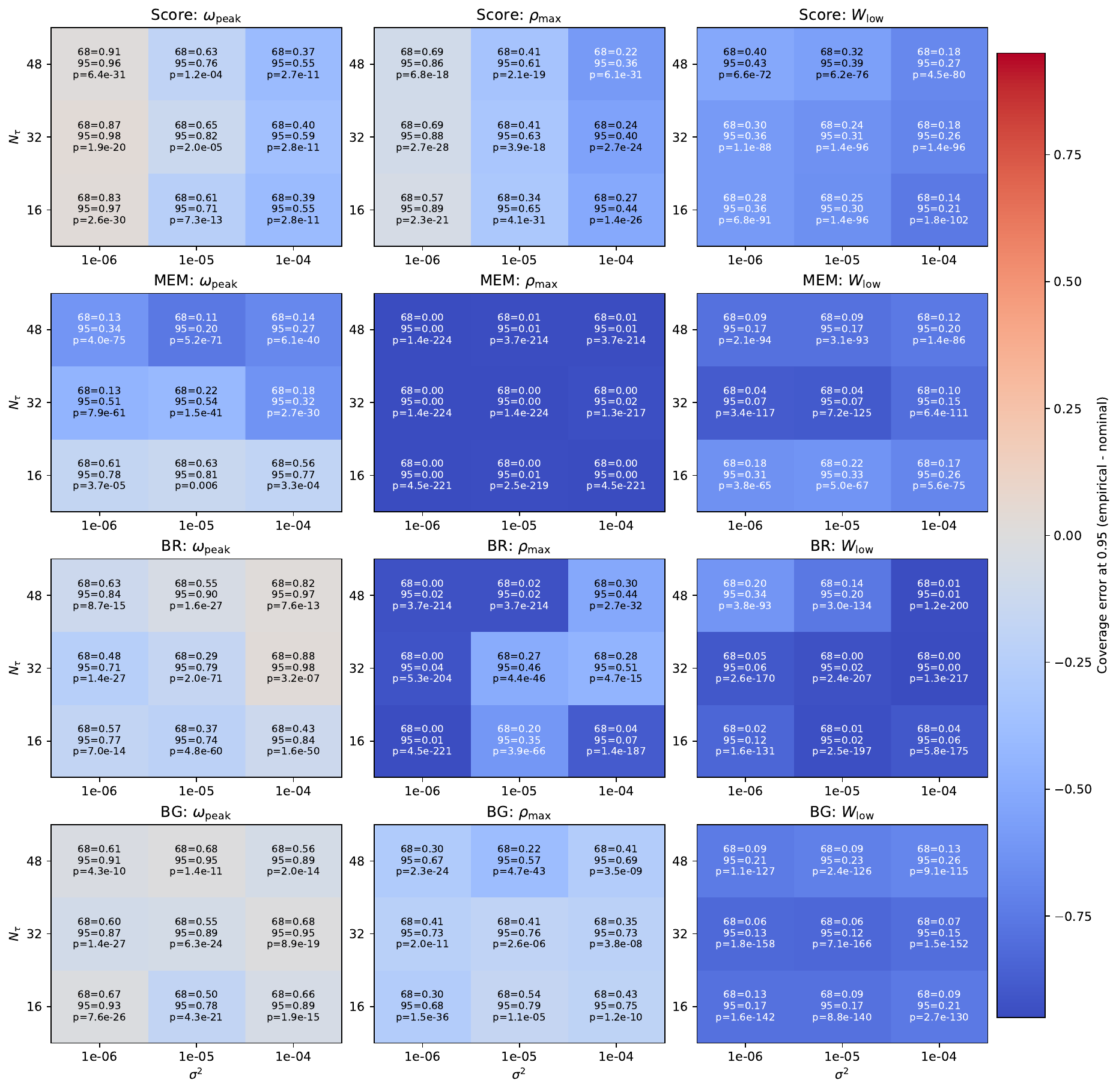}
  \caption{Full primary calibration stress matrix over \(\Ntau\) and \(\sigma^2\) at fixed Toeplitz correlation length \(\ell=0.25\).  Each cell reports the 68\% and 95\% empirical coverages and the rank-uniformity diagnostic used in the internal audit.  This figure is retained as the detailed diagnostic record; the main text uses the compressed coverage-range summary in Fig.~\ref{fig:stress_summary}.}
  \label{fig:stress_matrix_app}
\end{figure*}

This appendix records the 95\% coverage matrices for the primary zero-temperature stress test.  The grid is
\begin{equation}
  (\Ntau,\sigma^2)\in\{16,32,48\}\times\{10^{-6},10^{-5},10^{-4}\},
  \quad \ell=0.25.
  \label{eq:app_stress_grid}
\end{equation}
The tables below are the authoritative stress-matrix record.  Figure~\ref{fig:stress_matrix_app} shows the full heat-map visualization corresponding to the same numbers summarized in Fig.~\ref{fig:stress_summary}.

\begin{table}[t]
\caption{95\% empirical coverage for \(\wpeak\).}
\scriptsize
\label{tab:matrix_pp95_full}
\begin{ruledtabular}
\begin{tabular}{lcccc}
Setting \((\Ntau,\sigma^2)\) & Score & MEM-based & BG-type & BR-based\\
\hline
\((16,10^{-4})\) & 0.555 & 0.766 & 0.891 & 0.836\\
\((16,10^{-5})\) & 0.715 & 0.812 & 0.777 & 0.738\\
\((16,10^{-6})\) & 0.973 & 0.781 & 0.934 & 0.773\\
\((32,10^{-4})\) & 0.590 & 0.324 & 0.945 & 0.977\\
\((32,10^{-5})\) & 0.824 & 0.539 & 0.887 & 0.793\\
\((32,10^{-6})\) & 0.984 & 0.508 & 0.871 & 0.707\\
\((48,10^{-4})\) & 0.551 & 0.270 & 0.887 & 0.969\\
\((48,10^{-5})\) & 0.758 & 0.199 & 0.949 & 0.902\\
\((48,10^{-6})\) & 0.961 & 0.340 & 0.914 & 0.836
\end{tabular}
\end{ruledtabular}
\end{table}

\begin{table}[t]
\caption{95\% empirical coverage for \(\rpeak\).}
\scriptsize
\label{tab:matrix_ph95_full}
\begin{ruledtabular}
\begin{tabular}{lcccc}
Setting \((\Ntau,\sigma^2)\) & Score & MEM-based & BG-type & BR-based\\
\hline
\((16,10^{-4})\) & 0.441 & 0.004 & 0.754 & 0.070\\
\((16,10^{-5})\) & 0.648 & 0.012 & 0.793 & 0.348\\
\((16,10^{-6})\) & 0.891 & 0.004 & 0.684 & 0.008\\
\((32,10^{-4})\) & 0.402 & 0.016 & 0.730 & 0.508\\
\((32,10^{-5})\) & 0.633 & 0.000 & 0.762 & 0.457\\
\((32,10^{-6})\) & 0.875 & 0.000 & 0.727 & 0.035\\
\((48,10^{-4})\) & 0.359 & 0.008 & 0.688 & 0.438\\
\((48,10^{-5})\) & 0.609 & 0.012 & 0.570 & 0.023\\
\((48,10^{-6})\) & 0.863 & 0.000 & 0.668 & 0.023
\end{tabular}
\end{ruledtabular}
\end{table}

\begin{table}[t]
\caption{95\% empirical coverage for \(\Wlow=\int_0^3\rho(\omega)\dd\omega\).}
\scriptsize
\label{tab:matrix_lw95_full}
\begin{ruledtabular}
\begin{tabular}{lcccc}
Setting \((\Ntau,\sigma^2)\) & Score & MEM-based & BG-type & BR-based\\
\hline
\((16,10^{-4})\) & 0.207 & 0.262 & 0.207 & 0.059\\
\((16,10^{-5})\) & 0.305 & 0.328 & 0.168 & 0.020\\
\((16,10^{-6})\) & 0.359 & 0.312 & 0.168 & 0.117\\
\((32,10^{-4})\) & 0.258 & 0.148 & 0.152 & 0.000\\
\((32,10^{-5})\) & 0.312 & 0.066 & 0.125 & 0.023\\
\((32,10^{-6})\) & 0.359 & 0.070 & 0.133 & 0.059\\
\((48,10^{-4})\) & 0.273 & 0.203 & 0.262 & 0.008\\
\((48,10^{-5})\) & 0.387 & 0.168 & 0.230 & 0.199\\
\((48,10^{-6})\) & 0.434 & 0.168 & 0.211 & 0.340
\end{tabular}
\end{ruledtabular}
\end{table}

\begin{table*}[t]
\caption{Rank nonuniformity in the nearby-cutoff rescan for \(\Wlow(\omega_c)\).  Entries are \(D_{\rm KS}\) values of the SBC rank distribution for the same fixed representatives used in Table~\ref{tab:wlow_cutoff_rescan}.}
\scriptsize
\label{tab:wlow_cutoff_rescan_dks}
\begin{ruledtabular}
\begin{tabular}{lcccccc}
Method
& \multicolumn{3}{c}{\(P_{\rm mid}\)}
& \multicolumn{3}{c}{\(P_{\rm hard}\)}\\
& \(\omega_c=2\) & \(\omega_c=3\) & \(\omega_c=4\)
& \(\omega_c=2\) & \(\omega_c=3\) & \(\omega_c=4\)\\
\hline
Score     & 0.867 & 0.652 & 0.504 & 0.813 & 0.676 & 0.410\\
MEM-based & 0.824 & 0.742 & 0.653 & 0.758 & 0.574 & 0.349\\
BG-type   & 0.914 & 0.856 & 0.641 & 0.868 & 0.758 & 0.645\\
BR-based  & 0.984 & 0.957 & 0.906 & 0.969 & 0.879 & 0.820
\end{tabular}
\end{ruledtabular}
\end{table*}

The out-of-family score calibration test decomposes the total shift into peak-count, threshold, and tail components.  At the fiducial point the threshold-only shift is the leading degradation: \(\Wlow\) changes from \((0.777,0.840)\) to \((0.098,0.137)\) at the \((68\%,95\%)\) levels.  At the hard point it changes from \((0.449,0.594)\) to \((0.023,0.066)\).  This supports the interpretation that threshold support is the dominant domain-shift variable for the infrared weight in the present benchmark.

\section{Real-data shear-grid details}
\label{app:realdata_repro}

The real-data demonstration uses the viscosity-specific finite-temperature transfer setup rather than the generic zero-temperature scripts.  The mock source library uses
\begin{equation}
  \beta=1,
  \qquad \Ntau=37,
  \qquad K=1024,
  \qquad \omega_{\max}=20,
  \label{eq:visc_source_grid_app}
\end{equation}
with a fixed ultraviolet continuation used as a numerical convention in the transfer setup.  The observable-matching step slices the mock correlators to the 11 retained shear \(\tau T\) points, recomputes the normalization correlator on the same grid, and converts to \(y(\tau)=G(\tau)/G_{\rm norm}(\tau)\).  The finite-temperature \(\Wlow\) target in Eq.~\eqref{eq:real_wlow_target} uses the same dimensionless grid variable \(\Omega=\omega/T\) and cutoff \(\Omega_c=3\).  The published pointwise errors of Ref.~\cite{Altenkort:2023eav} define the diagonal covariance used for the standardized residual diagnostic.

The BG-style candidate grid is generated from fixed choices of the resolution regularization \(\lambda\), ridge stabilization, contact-jitter scale, oversampling factor, scale parameter, and ultraviolet cutoff convention.  The scan contains 8001 configurations.  The real-data \(\chi^2\) records include the configuration name, seed, unregularized standardized residual norm, and \(\chi^2/N_\tau\).  The best entry is configuration 07942, whose file name specifies the settings \(\lambda=10^{-3}\), ridge \(10^{-8}\), contact jitter \(10^{-14}\), oversampling 1, scale 3.5, and UV cutoff 10.  It has \(\chi^2/N_\tau=1.2668\).

The target-specific mock calibration is joined to the real \(\chi^2\) scan by configuration name.  For each joined reconstruction setting we compute
\begin{equation}
  J_W=\max_{\alpha\in\{0.68,0.95\}}|\widehat C_\alpha(\Wlow)-\alpha|
  +0.2D_{\rm KS}(\Wlow),
  \label{eq:real_jw_app}
\end{equation}
which is the quantity plotted in Fig.~\ref{fig:real_shear_chi2_calibration_map}.  No additional physical-diagnostic penalty is applied in this displayed shear-grid map after invalid settings have been removed.  Within the \(\chi^2/N_\tau<4\) compatibility set, the minimum stored value of \(J_W\) is shared by 20 configurations because the finite mock sample makes the coverage entries discrete; 07802 is used as a representative member of this minimum-\(J_W\) class.  The selected real-data \(\Wlow\) table is generated from stored samples for three representative configurations: 07802, 07942, and 07812.  Their 68\% widths are 3.114, 7.155, and 5.141, respectively, while their medians are all close to 6.3--6.5 in the working units.

The direct feasibility diagnostic is separate from the candidate-grid calibration test.  It solves a non-negative least-squares problem under the same preprocessing, diagonal covariance, finite-temperature kernel, working grid, and fixed ultraviolet convention,
\begin{equation}
 \widehat\rho_{\lambda}=\arg\min_{\rho_k\ge0}
 \left\|\Sig^{-1/2}(\widetilde K\rho+g_{\rm UV}-y)\right\|_2^2
 +\lambda\|L\rho\|_2^2,
 \label{eq:direct_feasibility_app}
\end{equation}
where \(L\) is a finite-difference matrix and \(g_{\rm UV}\) denotes the fixed ultraviolet contribution.  The quoted feasibility number is the unregularized data term evaluated at the fitted spectrum,
\begin{equation}
  \frac{\chi^2_{\rm data}}{N_\tau}
  =\frac{1}{N_\tau}
  \left\|\Sig^{-1/2}(\widetilde K\widehat\rho_\lambda+g_{\rm UV}-y)\right\|_2^2 .
  \label{eq:direct_feasibility_chi2_app}
\end{equation}
The best shear value is \(0.855\).  This feasibility diagnostic is not an uncertainty report, is not used to tune the scanned BG-style candidates, and is not a transport extraction.

The release includes the joined \(\chi^2\)-calibration table, selected \(\Wlow\) summaries, the peak-height coverage columns used in Fig.~\ref{fig:real_shear_chi2_calibration_map}, and per-time residual diagnostics needed to reproduce Figs.~\ref{fig:real_shear_chi2_calibration_map} and \ref{fig:real_shear_supporting_checks}.  For the sample-mean central spectra displayed in Fig.~\ref{fig:real_shear_supporting_checks}, the recomputed \(\chi^2/N_\tau\) values are 1.289, 1.472, and 1.326 for 07802, 07942, and 07812, with \(\max_i|z_i|=1.654\), 2.048, and 1.801.  The exact linear BG central spectra are also archived; they give \(\chi^2/N_\tau=1.372\), 1.377, and 1.372.  These per-time diagnostics are a visual compatibility check under explicit central-curve conventions, while the main text uses the archived scalar \(\chi^2\) scan for the candidate-grid compatibility criterion.

\bibliography{apssamp}

\end{document}